# Multi-objective Eco-Routing Model Development and Evaluation for Battery Electric Vehicles

**Kyoungho Ahn**
Center for Sustainable Mobility, Virginia Tech Transportation Institute
3500 Transportation Research Plaza, Blacksburg, VA 24061
Phone: (540) 231-1500 Fax: (540) 231-1555
kahn@vt.edu
ORCID number: 0000-0003-4272-3840

**Youssef Bichiou**
Center for Sustainable Mobility, Virginia Tech Transportation Institute
3500 Transportation Research Plaza, Blacksburg, VA 24061
Phone: (540) 231-1500 Fax: (540) 231-1555
ybichiou@vtti.vt.edu

**Mohamed Farag**
Center for Sustainable Mobility, Virginia Tech Transportation Institute
3500 Transportation Research Plaza, Blacksburg, VA 24061
Phone: (540) 231-1500 Fax: (540) 231-1555
College of Computing and Information Technology
Arab Academy for Science, Technology, and Maritime Transport
Alexandria, Egypt
mfarag@vtti.vt.edu

**Hesham A. Rakha**
Charles E. Via, Jr. Department of Civil and Environmental Engineering
Center for Sustainable Mobility, Virginia Tech Transportation Institute
3500 Transportation Research Plaza, Blacksburg, VA 24061
Phone: (540) 231-1505; Fax: (540) 231-1555
hrakha@vt.edu
ORCID number: 0000-0002-5845-2929

**ABSTRACT**

This paper develops and investigates the impacts of multi-objective Nash optimum (user equilibrium) traffic assignment on a large-scale network for battery electric vehicles (BEVs) and internal combustion engine vehicles (ICEVs) in a microscopic traffic simulation environment. Eco-routing is a technique that finds the most energy efficient route. ICEV and BEV energy consumption patterns are significantly different with regard to their sensitivity to driving cycles. Unlike ICEVs, BEVs are more energy efficient on low-speed arterial trips compared to highway trips. Different energy consumption patterns require different eco-routing strategies for ICEVs and BEVs. This study found that eco-routing could reduce energy consumption for BEVs but also significantly increases their average travel time. The simulation study found that multi-objective routing could reduce the energy consumption of BEVs by 13.5%, 14.2%, 12.9%, and 10.7%, as well as the fuel consumption of ICEVs by 0.1%, 4.3%, 3.4%, and 10.6% for "not congested, "slightly congested," "moderately congested," and "highly congested" conditions, respectively. The study also found that multi-objective user equilibrium routing reduced the average vehicle travel time by up to 10.1% compared to the standard user equilibrium traffic assignment for the highly congested conditions, producing a solution closer to the system optimum traffic assignment. The results indicate that the multi-objective eco-routing can effectively reduce fuel/energy consumption with minimum impacts on travel times for both BEVs and ICEVs.

**INTRODUCTION**

This study investigates and compares the impacts of different eco-routing strategies for battery electric vehicles (BEVs) and internal combustion engine vehicles (ICEVs) on a large-scale network. Passenger electric vehicle (EV) sales reached 2.1 million in 2019, up from 450,000 in 2015. A recent report predicted that 50% or more of all passenger vehicles sold would be EVs by 2040 [1]. We expect that over the next few years both more luxury and affordable EV models with longer electric range will be available and more EVs will occupy the roads.

Eco-routing is a technique that finds the most energy-efficient route. Traditionally, drivers determine their route to reduce delays based on their urgency, experience, and current information, including travel time, trip distance, and other trip-related factors. Drivers typically do not consider energy and environmental impacts in their routing decisions [2]. Eco-routing can improve vehicle energy efficiency and reduce vehicle emissions by determining the most energy- and eco-friendly route. Using eco-routing, drivers can reduce fuel/energy consumption and vehicle emissions. Further, due to limited battery capacities and driving ranges, eco-routing can be a valuable feature for BEVs, particularly when a charging area is not immediately available.

The fuel/energy consumption patterns of ICEVs and EVs differ significantly with regard to driving cycles. A previous study found that BEVs are more energy efficient on low-speed arterial trips compared to highway trips. For instance, the study demonstrated that fuel consumption rates on a Freeway G cycle and a Local cycle were very similar for a test ICEV, but that a BEV's energy consumption rates on these two cycles were significantly different. The study demonstrated that the difference was due to the energy recovered during braking, and that the BEV recovered a higher amount of energy on the Local cycle [3]. BEVs can recover more energy in urban arterial driving conditions due to increased braking, while ICEVs' fuel efficiency is higher during uninterrupted freeway operational conditions. The different energy consumption patterns of ICEVs, BEVs, and hybrid electric vehicles will require different eco-routing strategies. However, the effects of eco-routing for these three vehicle types have not been systematically investigated and compared.

Connected vehicle (CV) technologies allow vehicles and transportation infrastructure to communicate with each other to improve the safety, mobility, and energy efficiency of the transportation system. CV systems comprise sets of applications that connect vehicles to each other and to the road infrastructure using vehicle-to-vehicle (V2V) and vehicle-to-infrastructure (V2I) communications, collectively known as V2X. These new technological advancements have the potential to improve the efficiency and sustainability of our transportation system. Within a CV environment, vehicles can exchange data with roadside equipment, allowing environmental and traffic-related data along different roadway sections to be shared and used for routing purposes. The proposed study assumes that vehicles will use V2I communication to report their traffic and environmental parameters on individual links to roadside equipment or directly to a traffic management center (TMC) via wireless communication within the CV environment.

The objective of this study is to develop energy efficient routing strategies for BEVs that minimize their energy consumption while at the same time do not produce large increases in vehicular travel times. The remainder of the paper is organized as follows: the next section reports the state-of-the art eco-routing efforts for both ICEVs and BEVs. Then, the paper describes the BEV and ICEV fuel/energy consumption models that were utilized for this study. The next sections describe the eco-routing logic. Then the paper analyzes the various eco-routing options and demonstrates results of a simulation model. Finally, the study conclusions are presented.

**LITERATURE REVIEW**

A number of studies have been conducted to develop and evaluate eco-routing strategies. An earlier study by Ahn and Rakha [4] investigated the impacts of route choice decisions on vehicle energy consumption and emission rates for different vehicle types using microscopic and macroscopic emission estimation tools. The results demonstrated that the faster highway route is not always the ideal route choice from environmental and energy consumption perspectives. Specifically, the study found that significant improvements to energy and air quality could be achieved when motorists utilized a slower arterial route, even though they incurred additional travel time. The study also demonstrated that macroscopic emission estimation tools could produce erroneous conclusions given that they ignore transient vehicle behavior along a route. The findings suggest that emission- and energy-optimized traffic assignment can significantly improve emissions over the standard user equilibrium and system optimum assignment formulations. The researchers also developed a stochastic feedback eco-routing system that builds on vehicle connectivity and quantified the system-wide impacts for various levels of market penetration and congestion in downtown Cleveland and Columbus, Ohio. The findings indicated network-wide fuel consumption savings between 3.3% and 9.3% when compared to typical routing strategies that minimize travel time [5, 6]. The simulation results also demonstrated that eco-routing vehicles do not always reduce fuel consumption compared to non-eco-routing vehicles. In particular, the fuel savings for eco-routing vehicles are sensitive to the network configuration, congestion levels, and eco-routing vehicle market penetration conditions. Simulation results showed that, in general, when 90% and 100% of eco-routing vehicles are assigned, total system-wide fuel consumption is minimized in most cases.

Richter et al. (2012) presented a study to show the potential energy savings and differences between routes for an ICEV, BEV and plug-in hybrid electric vehicle (PHEV). The analysis was performed using the ULTraSim traffic simulator, which incorporated submicroscopic BEV, PHEV, and ICEV vehicle models. The study examined the fuel savings of the eco-route for each type of vehicle in comparison to the shortest route, finding that the savings potential was dependent on route planning. The study recommends considering the different drive trains in eco-route calculation [7].

Liu et al. (2014) presented a minimum-cost path optimization scenario for real-time pricing with multiple charging stops in long-distance origin-destination trips for BEVs. The study utilized dynamic programming to solve the optimum cost problem with a travel time limitation that considered charging control. Strongly connected component algorithms were designed to reduce the computational complexity. The Improved Chrono-SPT (shortest path tree) was designed to provide an optimal routing and charging policy. The simulation results proved the effectiveness of the proposed approach [8].

Artmeier et al. (2010) investigated an energy-efficient path for BEVs with recuperation in a graph-theoretical context, which extended a general shortest path problem. The study modeled energy-optimal routing as a shortest path problem with various constraints and also considered energy costs or gains that might result from speed variability cost with different cruise speeds. The developed model was implemented into an energy-efficient prototype navigation system [9].

Energy-optimal routing for BEVs was also investigated by Sachenbacher et al. (2011). Their study claims that standard routing does not work for EVs due to their use of regenerative braking, along with the complexity of a number of parameters, such as vehicle load and auxiliary usages, as well as battery capacity limitations. The study proposed an Energy A* search algorithm to overcome the challenges and shows how battery constraints can be dynamically incorporated into the algorithm. Experimental test results with real road networks found that the proposed method was effective and faster compared to the generic framework using the Dijkstra or Pallottino strategy [10].

Bhavsar et al. performed a study in which they developed an integration simulation tool, CUIntegration, to evaluate vehicle routing strategies' effects on energy consumption and other traffic-related measures for ICEVs and alternative fuel vehicles, including PHEVs and BEVs. CUIntegration incorporates a routing strategy developed using MATLAB with the VISSIM microscopic traffic simulation software. The simulation study found that energy optimization resulted in about 30% savings in the EV's energy consumption, and travel time optimization resulted in about 65% savings in travel time with increased overall energy consumption [11].

The impact of driver behavior on BEVs' energy consumption was investigated by Bingham et al. The study found that energy consumption could be significantly reduced by eliminating acceleration and deceleration behavior throughout the tested driving cycle. The study reports that good driving can reduce total energy consumption by 30% compared to more aggressive driving based on the specific cases analyzed. The study also recommended considering "hotel loads" (i.e., with active air-conditioning and heating) for evaluating energy efficiency. Further, the study suggests using appropriate traffic management techniques by reducing periods of transient acceleration/deceleration and promoting consistent speed levels to improve fuel economy [12].

Yi et al. [13] studied the energy impact for eco-routing of an EV fleet under different ambient temperatures. The study used a data-driven EV energy consumption model, which was developed using data from five Nissan Leaf taxis in New York City. The proposed data-driven energy consumption model utilized average vehicle speed and ambient temperature of the trip as input variables. The objective function was to determine a path from an origin to a destination to minimize the total energy cost by satisfying the travel time and final energy state requirements. The study explained that the developed eco-routing and charging decision-making framework could help construct sustainable infrastructure for EV fleet trip-level energy management in real-world applications.

Abousleiman et al. [14] developed an energy-efficient routing model for EVs using an ant-colony-based optimization technique. The study tested the proposed model using a 2013 Fiat 500e EV. The study utilized a power-based energy consumption model assuming a standard driving style and typical powertrain component parameters without using a dynamic regenerative braking estimation. In particular, the study utilized a fixed acceleration and deceleration rate of 1.78 m/s$^2$. The study found that the proposed eco-routing method was, on average, 9% more efficient than the routes that were suggested by Google Maps and MapQuest. Abousleiman and Rawashdeh also investigated various eco-routing algorithms for EVs using heuristic methods based on Particle Swarm Optimization and Tabu Search. The studies focused on simplified algorithms without proving the implementation with a real-world example [15-17].

Ma et al. analyzed the environmental costs of EVs on the traffic network using a revised stochastic user equilibrium model and a revised version of the method of successive averages. The study found that the existence of EVs affected drivers' route choice behavior and also reduced the generalized costs of both EV drivers and ICEV drivers [18].

Most of the eco-routing research on BEV routing relies on simple aggregate energy consumption models (e.g., average speed model) and/or testing on a simple network. These approaches may produce incorrect results and have not quantified their impacts on large urban networks. One of the major benefits of BEVs is their regenerative braking system, which allows for the recovery of energy while braking. Most studies assume a regenerative braking factor mainly dependent on the vehicle speed or an average regenerative braking energy efficiency and usually do not take these factors into consideration while routing BEVs. These approaches cannot accurately estimate and distinguish the BEV's energy consumption between facilities that operate at the same average speed. In order to overcome these issues, this study utilizes microscopic fuel/energy models to develop a multi-objective feedback routing system that considers the vehicle energy consumption and tests the proposed system on a large urban network. Specifically, the study utilizes a microscopic BEV energy consumption model that can capture instantaneous braking energy regeneration, which is not available in most BEV energy models, to compute the link cost function and route BEVs in an energy-optimum manner.

**ENERGY CONSUMPTION MODELS**

This section describes the fuel and energy consumption models for ICEVs and BEVs used for the eco-routing study. The VT-Micro model was utilized to estimate the vehicle fuel consumption level for ICEVs in the simulation runs. The VT-Micro model is a mathematical model that estimates vehicle fuel consumption and emission levels for individual and/or composite vehicles using instantaneous speed and acceleration as explanatory variables. The model utilizes a number of data sources, including data collected at the Oak Ridge National Laboratory (ORNL) (nine vehicles) and the Environmental Protection Agency (101 vehicles). In this study, ORNL vehicles were utilized for the analysis.

The VT-Micro model was developed as a regression model resulting from experimentation with numerous polynomial combinations of speed and acceleration to construct a dual-regime model of the following form:

$$F(t) = \begin{cases} \exp\left(\sum_{i=0}^{3}\sum_{j=0}^{3} L_{i,j} u(t)^i a(t)^j\right) & \forall\, a(t) \geq 0 \\ \exp\left(\sum_{i=0}^{3}\sum_{j=0}^{3} M_{i,j} u(t)^i a(t)^j\right) & \forall\, a(t) < 0 \end{cases} \quad (1)$$

where $L^e_{i,j}$ and $M^e_{i,j}$ represent regression coefficients for measure of effectiveness (measures of effectiveness or $e$) at speed exponent $i$ and acceleration exponent $j$; $u$ is instantaneous speed; and $a$ is instantaneous acceleration rate. It should be noted that the intercept at zero speed and zero acceleration was estimated using the positive acceleration model and was fixed to ensure a continuous function between the two regression regimes. Consequently, the calibration of the model involves estimating a total of 32 parameters for each measure of effectiveness. The VT-Micro model fuel consumption and emission rates were found to be highly accurate compared to the original data, with coefficients of determination ($R^2$) ranging from 0.92 to 0.99. The model is easy to use for the evaluation of the environmental impacts of operational-level projects, including intelligent transport systems. A more detailed description of the model derivation is provided in the literature [19, 20].

The Virginia Tech Comprehensive Power-based EV Energy consumption Model (VT-CPEM) was selected to estimate BEV energy consumption levels [3]. The VT-CPEM model is a microscopic, power-based BEV energy model developed to estimate the instantaneous energy consumption of BEVs. The model uses instantaneous speed, acceleration, and grade information as input variables. The outputs of the model are the energy consumption (kWh/km), the instantaneous power consumed (kW), the instantaneous energy regenerated (kW), and the final state of charge of the electric battery (%). VT-CPEM has a simple structure that allows it to be implemented into other modeling tools, including microscopic traffic simulation models and in-vehicle/smartphone applications for real-time eco-driving and eco-routing. One of the major advantages of VT-CPFM is that it can capture instantaneous braking energy regeneration, which is not available in most EV energy models. The majority of EV energy models use an average regenerative braking energy efficiency [21] or a regenerative braking factor that depends mainly on the vehicle speed [22-24], neither of which can accurately capture the instantaneous regenerative braking energy, which was required in this study. VT-CPEM was validated against experimental data and found to accurately estimate energy consumption, producing an average error of 5.9% relative to empirical data.

**ECO-ROUTING LOGIC AND SIMULATION TEST**
We used a stochastic incremental feedback Nash optimum (user equilibrium) traffic assignment technique as the default eco-routing model. The routing is termed ECO-Subpopulation Feedback Assignment (ECO-SFA). A vehicle route is selected based on the fuel/energy consumption experiences of other vehicles within the same vehicle class. All drivers within a specific class are divided into five subpopulations, each comprising 20% of all drivers. The paths for each of these subpopulations are then updated every $t$ seconds during the simulation based on real-time measurements of link fuel/energy consumption levels for the specific vehicle class under consideration. The minimum path updates of each vehicle subpopulation are staggered in time to avoid situations in which all vehicles select identical paths if the minimum and next minimum path are similar. This logic results in 20% of the driver paths being updated every $t/5$ s. In this study, we updated the results every 60 s.

The selection of the next link is decided using a vehicle-specific array that lists the entire sequence of links from a vehicle's current link to its destination. Upon the completion of any link, a vehicle simply submits its fuel consumption on the link and then queries this array to determine which link it should utilize next to reach its ultimate destination in the most efficient manner depending on the objective function (e.g. travel time, fuel consumption, energy consumption, or multi-objective). The vehicle only uses the experiences of other vehicles in the same class to update the link cost function estimates on a link. This allows for a multi-class, user equilibrium, stochastic, dynamic (based on the update frequency specified by the user) eco-routing system. The routing is multi-class because vehicles are only affected by experiences of other vehicles in the same class. The assignment is the Nash optimum (user equilibrium) because each vehicle attempts to optimize its cost function. The routing is stochastic because the system can introduce white noise to the link-specific cost function. The user specifies the coefficient of variation of the white noise distribution. Furthermore, the routing is dynamic because vehicles can change their routes en-route using the latest information provided by other vehicles traversing the network.

The routing could be implemented in the field through the use of V2I or V2V communication, where vehicles would report their fuel/energy consumption levels on individual links to roadside units [25]. The information would then be sent to a traffic management center (TMC), where it would be fused with input from other vehicles traversing the network. Routing decisions would be decentralized, with individual vehicles making routing decisions based on the latest information received from the TMC.

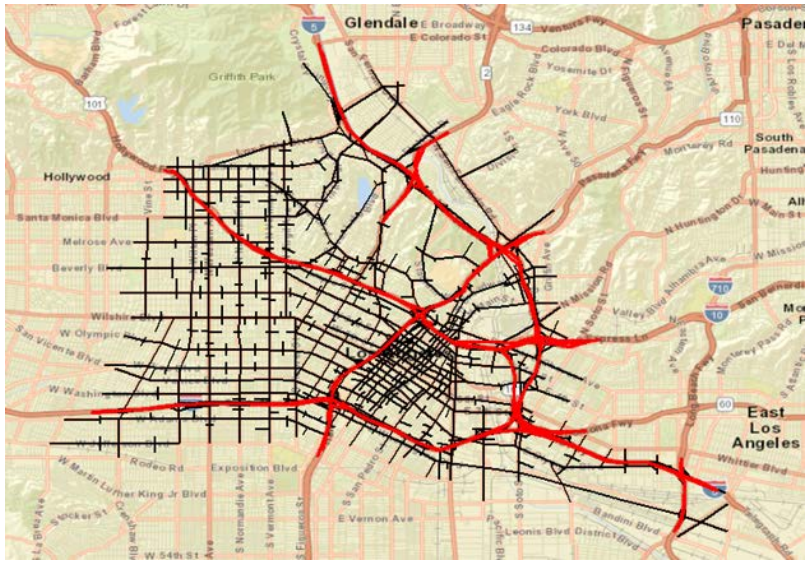

**Figure 1 Los Angeles test network**

The study tested the impacts of eco-routing using a real-world network in Los Angeles (LA), California, as shown in Figure 1. This network was calibrated to the local conditions in LA. There are several advantages to using real-world networks. First, both networks reflect real-world traffic conditions in large metropolitan areas under typical peak demand levels. Second, each network contains a range of road types, including multiple interstate highways, highway ramps, major arterials, and local connectors. Third, the network includes traffic control infrastructures, such as stop signs and traffic signals using optimized signal timing plans. Actual traffic signal timing data were coded in the simulation model. The LA test network includes interstates I-5, I-10, I-710, I-110, and U.S. route 101, which serve the downtown area. The network is composed of more than 3,500 links with a traffic demand of roughly 140,000 vehicles per hour during the morning peak hour. The LA network was constructed using 1,624 nodes, 3,556 links, 457 traffic signals, and 81,858 origin-demand pairs. The traffic demand was estimated using the QUEENSOD software [26], which computes the most-likely time dependent static traffic assignment and OD demand by iteratively minimizing the error between the observed traffic counts obtained from selected loop detectors and the corresponding estimated traffic volume. Dynamic OD demands were then estimated using an iterative procedure described in [27]. The estimated OD matrix provided a good match to the field observed traffic counts with an $R^2$ of 0.9.

Figure 2 demonstrates the impacts of the eco-routing system for BEVs. The study utilized a 2015 Nissan Leaf to represent BEVs. We compared three routing options, including a traditional travel time (TT) routing (or also known as a user equilibrium assignment method), a TT feedback routing that uses a subpopulation feedback assignment based on link travel times, and an eco-feedback routing. We utilized the INTEGRATION traffic simulation and assignment software to evaluate the impacts of various routing options. In the study, all vehicles on the network utilized TT-routing, TT-feedback routing, or eco-feedback routing with four different congestion levels. We tested "not congested," "slightly congested," "moderately congested," and "highly congested conditions," which represent 25%, 50%, 75%, and 100% of weekday morning peak traffic demands, respectively.

The figure demonstrates that eco-driving reduces the energy consumption for BEVs compared to both TT-routing and TT-feedback routing methods. Also, the figure shows that the TT-feedback routing is slight more energy efficient than the TT-routing, reducing energy consumption by up to 1.9%. The results indicate that the eco-routing saved energy by 28.1%, 24.2%, 21.4%, and 17.4% for "not congested," "slightly congested," "moderately congested," and "highly congested" conditions, respectively. The figure also shows the travel time impacts of three routing options. The results indicate that eco-routing significantly increases vehicle travel time up to 285.5% compared to the TT-routing option. In particular, we found that the average travel time was increased from 1077.1 s to 3240.7 s when all drivers choose eco-routing. Due to the increased travel time, the 100% eco-routing option is not a good choice for most drivers even though they can reduce energy consumption by up to 28.1%.

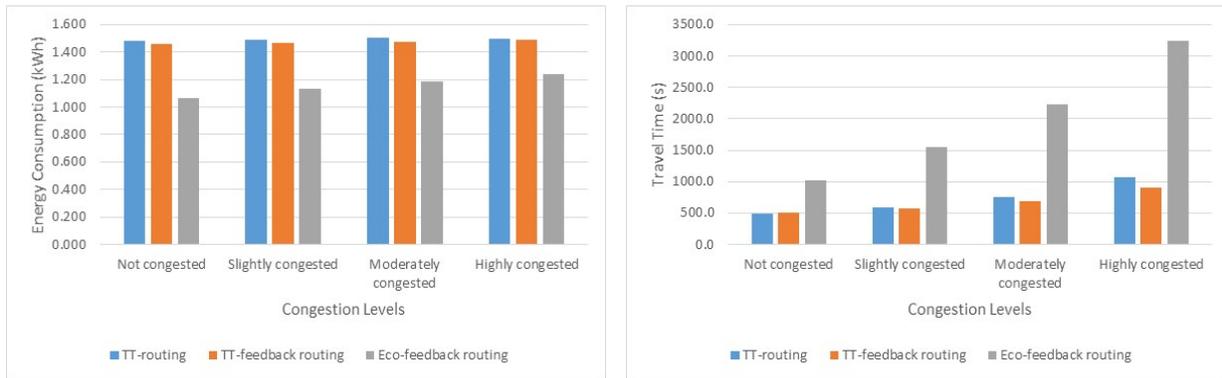

(a) BEV – Energy consumption  (b) BEV – Travel time

**Figure 2 Energy consumption and travel for BEV eco-routing**

## MULTI-OBJECTIVE ECO-ROUTING MODEL

We developed a multi-objective eco-routing model to reduce both travel time and fuel/energy consumption. The model introduced a link cost function that uses the specification of driver value of time and the cost of fuel/energy on specific links as shown in (1).

$$Link\_cost_l = (1 - a) \times TT_l + a \times energy\_consump_l \times \frac{cost\_factor}{value\_time} \qquad (1)$$

where $TT_i$ is a travel time (s) on link $i$; $a$ is a multi-objective coefficient; $energy\_consump_i$ is the fuel/energy consumption (l or kwh) on link $i$; $cost\_factor$ is the cost of fuel/energy; and $value\_time$ is the value of time. Vehicles select a next link based on the link cost instead of using either a travel time or fuel/energy consumption data. Similar to the other feedback routing options, vehicles only use the results of other vehicles in the same class to update the link cost estimates on a link. We used $10 for the value of time and $0.1319/kWh for electricity rate based on average electricity rate in the United States [28] for this study. Table 1 summarized the routing options utilized in this study.

**Table 1 Routing Options**

| Routing Option | Description |
|---|---|
| TT-routing | Travel time or user equilibrium (UE) traffic assignment using the Frank-Wolfe algorithm if the route travel times are deterministic or the method of successive averages if errors are introduced to the link travel times - the routing is selected based on the vehicle travel time. In our case we used the Frank-Wolfe algorithm to solve the UE problem. |
| TT-feedback routing | Travel time feedback UE traffic assignment - the routing is selected based on the travel time experiences of other vehicles within the same vehicle class. |
| Eco-routing | Eco-feedback UE traffic assignment - the routing is selected based on the fuel/energy consumption experiences of other vehicles within the same vehicle class. |
| MO-routing 1 | Multi-objective UE traffic assignment – the routing is selected based on the link cost function that is based on both the travel time and fuel/energy consumption. MO-routing 1 utilized 0.01 as the multi-objective coefficient. |
| MO-routing 2 | Multi-objective UE traffic assignment – the routing is selected based on the link cost function that is based on both the travel time and fuel/energy consumption. MO-routing 1 utilized 0.05 as the multi-objective coefficient. |

We tested the multi-objective eco-routing model using INTEGRATION and found that multi-objective eco-routing significantly reduces the energy consumption compared to the TT-routing and marginally increases the travel time compared to the travel time optimum routing, as shown in Figure 3. In particular, we used two multi-objective routings, MO-routing 1 and MO-routing 2, with coefficients 0.01 and 0.05, respectively, in this study. The absolute values of the multi-objective coefficients should be selected based on the cost factors and the values of time. We found that both MO-routing 1 and MO-routing 2 were effective compared to the TT-routing and the eco-routing in this case study. Specifically, MO-routing 1 was more effective on travel time, while MO-routing 2 saved more energy. We found that MO-routing 1 reduced BEV energy consumption by 24.3%, 21.0%, 18.8%, and 15.1% compared to the TT-routing and reduced travel time by 24.9%, 43.6%, 49.8%, and 54.7% compared to the eco-routing for "not congested," "slightly congested," "moderately congested," and "highly congested" conditions, respectively. However, we found that MO-routing 1 still increased the travel time up to 53% compared to the TT-routing.

The figure also shows the average travel distance for different routing options and congestion levels. We found that the TT-routing vehicles traveled longer trips than the eco-routing vehicles and the MO-routing 1 vehicles utilized the shortest distance routes. The study found that the MO-routing 1 vehicles saved trip distance by 2.7%, 2.8%, 2.4%, and 1.5% compared to the TT-routing for different congestion levels.

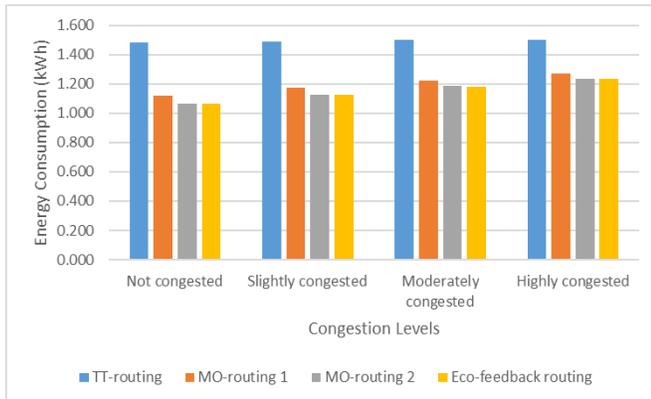
(a) BEV – Energy consumption

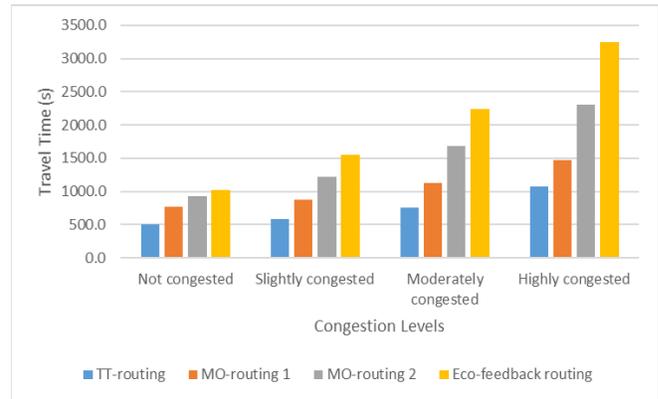
(b) BEV – Travel time

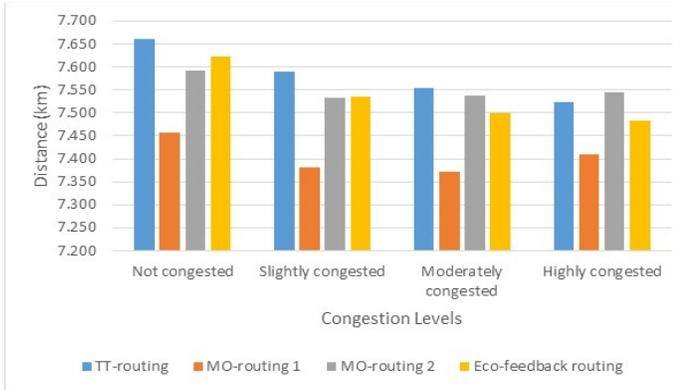

(c) BEV – Trip distance

**Figure 3 BEV energy consumption, travel time, and trip distance for different routing options**

The study also tested and compared two routing options in the LA network and investigated the impacts of different routing options. In particular, we tested 100% TT-routing, 50% TT-routing and 50% MO-routing 1, 50% TT-routing and 50% MO-routing 2, and 50% TT-routing and 50% eco-routing. For instance, 50% TT-routing and 50% MO-routing 1 indicates that 50% of vehicles on the network utilize MO-routing 1 method and the other 50% of vehicles use the TT-routing method.

As shown in Figure 4, the network performance is significantly improved when the vehicles choose two routing options in a network compared to a single routing option. In particular, the MO-routing 1 option reduces the energy consumption up to 26% and increases the travel time as little as 5.1% compared to the 100% TT-routing option. Compared to the single routing option, we found that the energy savings are comparable but the MO-routing 1 option considerably reduced the average vehicle travel time in this case study. The study also found that the MO-routing 1 option reduced the average vehicle trip distance up to 1.7%.

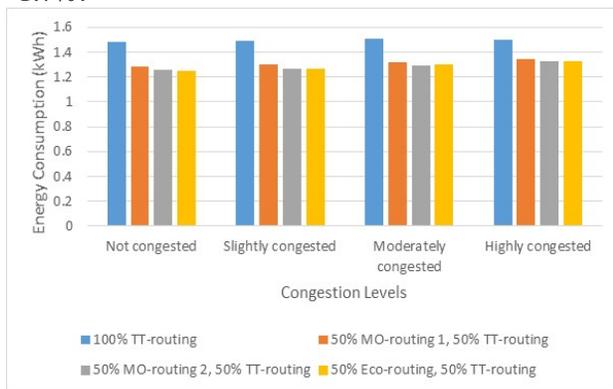

(a) BEV – Energy consumption

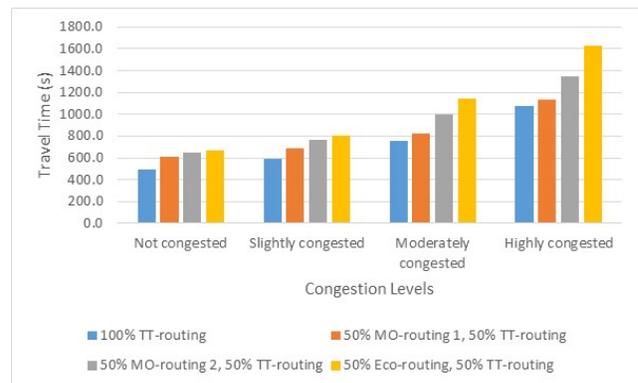

(b) BEV – Travel time

Ahn, Bichiou, Farag, and Rakha                                                                                                                   10

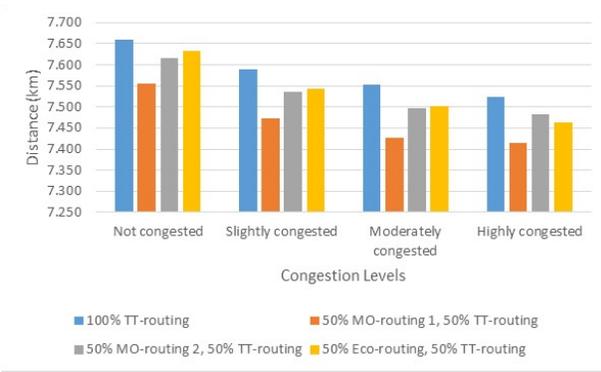

(c) BEV – Trip distance

**Figure 4 Energy consumption, travel time, and trip distance for BEV 50% eco-routing**

Figure 5 compares the energy consumption and the average travel time of BEVs for 50% TT-routing and 50% MO-routing 1 in the LA network. The figure shows that the MO-routing reduces energy by 25.4%, 25.7%, 24%, and 22.8%. The TT-routing vehicle reduces the average travel time by 48.5%, 43.6%, 32.5%, and 21.3% for "not congested," "slightly congested," "moderately congested," and "highly congested" conditions, respectively. The results indicate that drivers have good options for either a fast trip or an energy-efficient trip when two routing options are available.

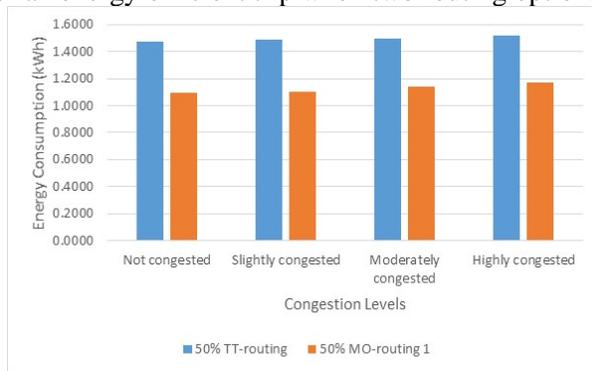 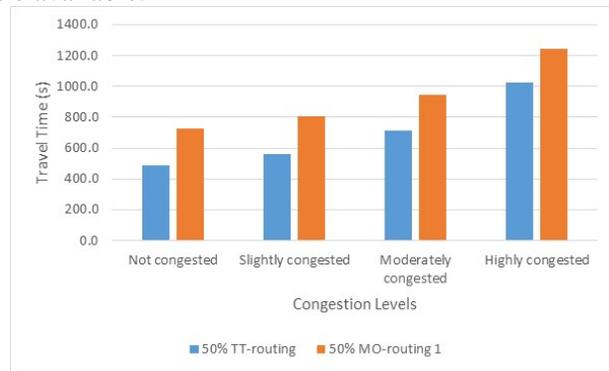

(a) Energy consumption                                                      (b) Travel time

**Figure 5 Comparison of energy consumption and travel time for BEV multi-objective routing**

The study also investigated the impacts of the eco-routing options for ICEVs and BEVs in the LA network, as illustrated in Figure 6. We assigned four vehicle classes, including two 25% ICEVs and two 25% BEVs, into the LA network. For instance, MO routing 1 represents 25% ICEV TT-feedback routing, 25% ICEV MO-routing 1, 25% BEV TT-feedback routing, and 25% BEV MO-routing 1. Figure 6 shows that MO-routing options reduce the energy consumption of BEVs and the fuel consumption of ICEVs, and also reduce the travel time compared to the TT-routing. For this specific case study, we found that MO-routing 1 reduced BEV energy consumption by 13.5%, 14.2%, 12.9%, and 10.7%, ICEV fuel consumption by 0.1%, 4.3%, 3.4%, and 10.6% for "not congested," "slightly congested," "moderately congested," and "highly congested" conditions, respectively. The study also found that MO-routing 1 reduced the average vehicle travel time by 0.4% and 10.1% for "moderately congested" and "highly congested" conditions, respectively, and slightly increased the travel times for less-congested conditions. The results indicate that multi-class eco-routing options can reduce both fuel/energy consumption and travel time since multiple vehicle classes can find their own routes without congesting preferred links. The figure also illustrates the average travel distance for different routing options. The study found that the eco-routing option used the shortest distance trips for "not congested," "slightly congested," and "moderately congested" conditions, while the TT-routing option chose the shortest distance trips for "highly congested" condition.

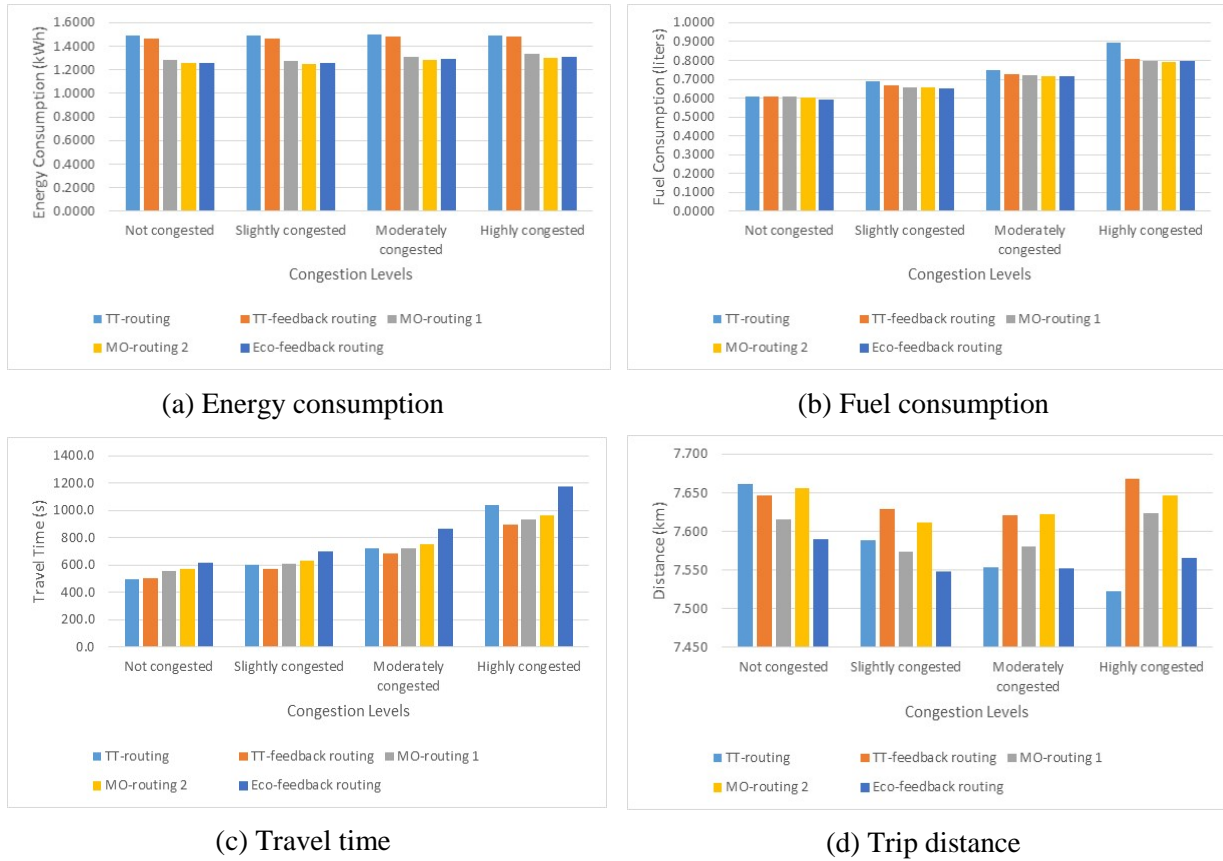

(a) Energy consumption  (b) Fuel consumption

(c) Travel time  (d) Trip distance

**Figure 6 Energy consumption, fuel consumption, travel time, and trip distance of different routing options for multi-class vehicle types**

Figure 7 compares the fuel/energy consumption and the travel time of ICEVs and BEVs for MO-routing 1 and the eco-routing options. The figure shows the simulation results of the highly congested condition. The study also demonstrates that MO-routing options outperform the eco-routing. The study found that both the MO-routing 1 and eco-routing vehicles consumed very similar amounts of fuel/energy except for the 25% BEV eco-routing vehicles, which reduced energy consumption by 5.4% compared to the MO routing 1 vehicles. However, the study found that the eco-routing vehicles increased the travel time up to 25.9% and 29.4% for ICEVs and BEVs. The results indicate that the MO-routing method can effectively reduce both fuel/energy consumption and travel time for different vehicle types.

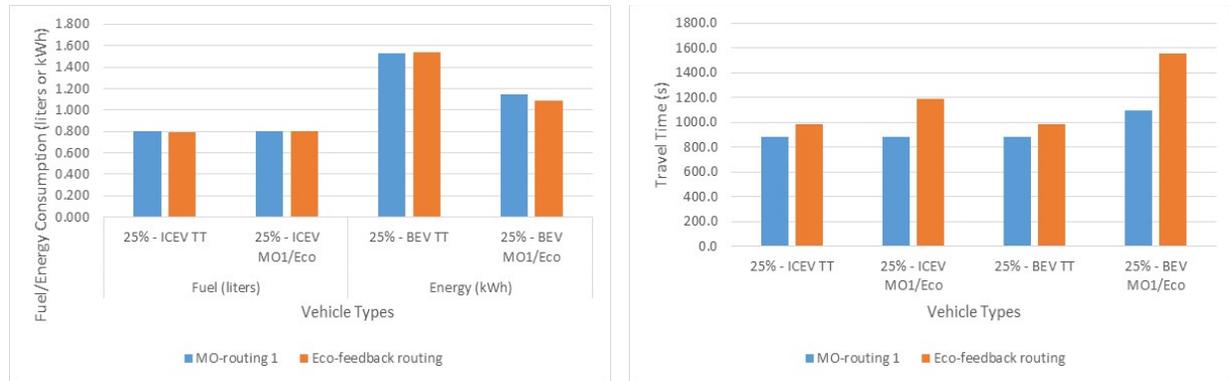

(a) Highly congested – Fuel/Energy consumption      (b) Highly congested – Travel time

**Figure 7 Fuel/energy consumption and travel time for high -congestion, multi-class routing options**

## CONCLUSIONS

This paper developed a BEV and ICEV multi-objective, multi-class, stochastic, user equilibrium traffic assignment model and quantified its performance on a large-scale network. The study found that the a single-objective energy efficient UE traffic assignment could reduce BEV energy consumption but also significantly increase their average travel time. The simulation study found that multi-objective eco-routing could reduce the energy consumption of BEVs by up to 14.2% and the fuel consumption of ICEVs up to 10.6% for different congestion levels. The study also found that multi-objective eco-routing reduced the average vehicle travel time by up to 10.1% relative to the travel time user equilibrium traffic assignment for the highly congested traffic conditions. The results indicate that the developed multi-objective eco-routing algorithm can reduce BEV and ICEV energy/fuel consumption with minimum impacts on travel times.

As with the case of any study, further research is warranted. Specifically, we recommend a further study that investigates the optimum coefficient for the multi-objective eco-routing model that can reduce fuel/energy consumption and travel time considering different combinations of various vehicle types, eco-routing market penetrations, and different congestion levels.

## ACKNOWLEDGEMENTS

This work was funded by the Department of Energy through the Office of Energy Efficiency and Renewable Energy (EERE), Vehicle Technologies Office, Energy Efficient Mobility Systems Program under award number DE-EE0008209.

## AUTHOR CONTRIBUTION STATEMENT

The authors confirm contributions to the paper as follows: study conception and design: Ahn, Rakha; multi-objective eco-routing model development: Bichiou, Farag, Rakha; simulation model analysis and interpretation of results: Ahn, Rakha; manuscript preparation: Ahn, Rakha. All authors reviewed the results and approved the final version of the manuscript.